\documentclass[prd, twocolumn, superscriptaddress, nofootinbib]{revtex4}
\usepackage{graphicx, epsfig}
\usepackage{amsmath}
\usepackage{amsfonts}

\def\hmpcinv{h\,{\rm Mpc}^{-1}}
\def\be{\begin{equation}}
\def\ee{\end{equation}}
\def\bea{\begin{eqnarray}}
\def\eea{\end{eqnarray}}
\newcommand{\zmax}{z_{\rm max}}
\newcommand{\ode}{\Omega_{DE}}
\newcommand{\tq}{\textbf{q}}
\newcommand{\tp}{\textbf{p}}
\newcommand{\tC}{\textbf{C}}
\newcommand{\tD}{\textbf{D}}
\newcommand{\tF}{\textbf{F}}
\newcommand{\tO}{\textbf{O}}
\newcommand{\tW}{\textbf{W}}
\newcommand{\tI}{\textbf{I}}

\begin{document}

\title{High-resolution temporal constraints on the dynamics of dark energy}

\author{Gong-Bo Zhao}
\affiliation{Department of Physics, Simon Fraser University,
Burnaby, BC, V5A 1S6, Canada}

\affiliation{Institute of High Energy Physics, Chinese Academy of
Science, P.O. Box 918-4, Beijing 100049, P. R. China}

\author{Dragan Huterer}
\affiliation{Department of Physics, University of Michigan, 450
Church St., Ann Arbor, MI, 48109}

\author{Xinmin Zhang}
\affiliation{Institute of High Energy Physics, Chinese Academy of
Science, P.O. Box 918-4, Beijing 100049, P. R. China}

\begin{abstract}
  We use the recent type Ia supernova, cosmic microwave background and
  large-scale structure data to shed light on the temporal evolution of the dark
  energy equation of state $w(z)$ out to redshift one.  We constrain the most
  flexible parametrization of dark energy to date, and include the
  dark energy perturbations consistently throughout. Interpreting our results
  via the principal component analysis, we find no significant evidence for
  dynamical dark energy: the cosmological constant model is consistent with data
  everywhere between redshift zero and one at 95$\%$ C.L.
\end{abstract}

\maketitle

\emph{Introduction.} It has been recognized for some time now that
accurate reconstruction of the expansion history of the universe is
a crucial step towards understanding the physical mechanism behind
the accelerating universe. Early work on constraining the expansion
history had concentrated on parametrizing the equation of state
(ratio of pressure to density) of dark energy $w(z)$ via one or two
parameters and measuring them together with the energy density
relative to critical $\ode$
\cite{paraw0,w0wa_data,paraw1,paraw2,relatedwork}.  More recently,
the program of reconstructing the expansion history had been
generalized to adding more parameters \cite{morepar_data} and
decorrelating them \cite{Huterer:2004ch,Wang:2005yaa,sn182}.
Nevertheless, the data at the time only allowed only up to three or
four band-powers of $w(z)$ to be considered, leading to ``jagged''
expansion history, especially in the range $0.5\lesssim z\lesssim
1$. In the future, a number of ``principal components'' of dark
energy --- parameters forming a natural basis in which the function
$w(z)$ can be expanded
--- will be measured \cite{PC,dePutter:2007kf}.

In this paper, we utilize a new variant of the nearly model-independent
approach to reconstruct $w(z)$ of dark energy with the latest astronomical
observations including type Ia supernovae (SNe Ia), power spectra of the
cosmic microwave background (CMB) anisotropies and galaxy distribution from
the large-scale structure (LSS).  We perform a full likelihood analysis using
the Markov chain Monte Carlo approach \cite{cosmomc}, and make sure to
properly take into account the dark energy perturbations \cite{Zhao:2005vj}.
Compared to previous work, our parametrization of the expansion history is
more flexible and, as we argue, robust and easy to implement.

\emph{Method and Data.} We consider the following cosmological
parameter set
\be \label{parameter}
      \{w_i,\omega_{b}, \omega_{c},
      \Theta_{s}, \tau,  n_{s}, \log[10^{10}A_{s}] \}~,
\ee
where $\omega_{b}\equiv\Omega_{b}h^{2}$ and $\omega_{c}\equiv\Omega_{c}h^{2}$
are the physical baryon and cold dark matter densities relative to critical,
$\Theta_{s}$ is $100\times$ the ratio of the sound horizon to the angular
diameter distance at decoupling, $\tau$ is the optical depth to re-ionization,
$A_{s}$ and $n_s$ are the amplitude of the primordial spectrum and the
spectral index respectively.

The remaining parameters, $w_i\,(i=1,2,..n)$, are the equation of state values
at $n$ specific fitting nodes $\{z_i\}$ --- redshifts at which we vary these
parameters independently.  We use a cubic spline to interpolate between these
nodes and obtain the full $w(z)$ in the redshift range $[0,\zmax]$.  We set
$\zmax=1.0$ since the number and quality of SN Ia data, and hence
constraints on dark energy, dramatically weaken beyond this redshift. At
$z>\zmax$, we assume $w(z)=-1$.  We have also tried allowing
$w(\zmax\leq z\leq z_{CMB})$ to be a single new parameter free to vary; we
found that the Markov chains do not converge because $w(z>\zmax)$ is
difficult to constrain using current data. Comparing this approach to setting
this parameter as $-1$, we found that the best $\chi^2$ remains nearly
unchanged.  Thus setting $w(z>\zmax)=-1$ is safe in the calculation. Overall,
our parametrization of $w(z)$ can be summarized as:
\be \label{paraw} w(z)= \left\{
    \begin{array}{ll}
          -1,&  \hbox{$z> \zmax$;} \\
      \hbox{  $w_i$}, & \hbox{$z\leq \zmax, z\in \{z_i\}$;} \\
      \hbox{ spline}, & \hbox{$z\leq \zmax, z\notin \{z_i\}$.}
    \end{array}
\right. \ee
%~~~~~~~~~~~~~~~~~(2)
For the basic results, we choose $n=6$ equation of state parameters
(we discuss later the sensitivity to varying $n$).

\begin{figure*}[htbp]
\begin{center}
\includegraphics[scale=0.5]{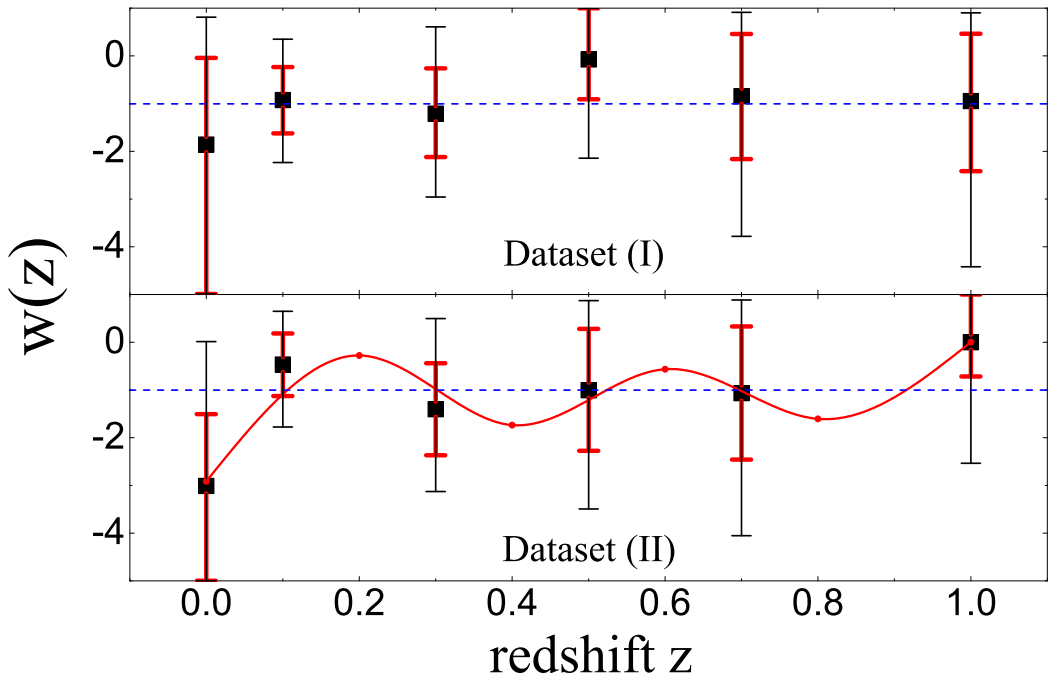}
\caption{Reconstruction of $w(z)$ using the datasets (I) (top panel)
and (II) (bottom panel); see text for details. The red and black
error bars correspond to 68 and 95$\%$ error bars respectively.
$\Lambda$CDM model is shown with the blue dashed line. In the bottom
panel, we also plot the reconstruction result with an alternative
placement of fitting nodes (red points and curve; see the text for
details).\label{fig1}}
\end{center}
\end{figure*}

It is important to take account of dark energy perturbations in the
analysis, otherwise the results may be biased
\cite{Weller:2003hw,paraw1}. In the conformal Newtonian gauge, one
can derive the following perturbation equations \cite{ma}:
\begin{eqnarray}
\label{eq:perturb1}
\dot\delta&=&-(1+w)(\theta-3\dot\Phi)
    -3\mathcal{H}(c_s^2-w)\delta \\
\label{eq:perturb2}
\dot\theta&=&-\mathcal{H}(1-3w)\theta-\frac{\dot{w}}{1+w}\theta
    +k^{2}(\frac{c_s^2}{{1+w}}\delta+ \Psi)
\end{eqnarray}
where the over-dot represents the derivative with respect to
conformal time,  ${c}_{s}^2\equiv\delta P/\delta\rho$ is the sound
speed, $\Psi$ is the Newtonian potential, $\mathcal{H}$ is the
(conformal time) Hubble parameter, $\delta$ is the density
perturbation and $\theta$ the velocity perturbation. One easily sees
that the perturbations $\delta$ and $\theta$ are divergent when
$w(z)$ crosses $-1$ even if nothing in the physical model should
diverge. One way to solve this problem is based on the quintom model
\cite{quintom} which we adopt here. In particular we introduce a
small positive constant $\epsilon$ to divide the full range of the
allowed values of $w$ into three regions: I) $ w > -1 + \epsilon$;
II) $-1 + \epsilon \geq w \geq-1 - \epsilon$; and III) $w < -1
-\epsilon $.  Neglecting the entropy perturbation contributions, for
the regions I) and III) the equation of state does not cross $-1$
and the perturbations are well defined by solving
Eq.~(\ref{eq:perturb1}) and (\ref{eq:perturb2}).  For the region
II), the density perturbation $\delta$, the velocity perturbation
$\theta$, and their derivatives are finite and continuous for the
realistic quintom dark energy models; therefore we set matching
conditions in region (II), $\dot\delta=0$ and $\dot\theta=0$.  This
is an approximate method to calculate DE perturbation during
crossing regime without introducing more
parameters\footnote{Ref\cite{nogo} has proved that in FRW
  cosmology, the equation of state of a single fluid or a single scalar field
  cannot cross $-1$. To realize such crossing, one has to introduce at least
  one more degree of freedom.}. The error in this approximation is
controllable and we have tested it in our numerical calculations; with
$\epsilon\sim10^{-5}$ we find that our method is a very good approximation to
the two-field quintom dark energy model. For more details of this method we
refer the reader to Refs.~\cite{paraw1,Zhao:2005vj}.

In our calculations we take the total likelihood as the products of
the separate likelihoods (${\bf \cal{L}}_i$) of CMB, LSS and SN~Ia,
i.e. defining $\chi_{i}^2 \equiv -2 \log {\bf \cal{L}}_i$, we use
\begin{equation}
\chi^2_{\rm total} = \chi^2_{\rm CMB} + \chi^2_{\rm LSS} +
\chi^2_{\rm SNIa}~.\label{chi2}
\end{equation}
For the CMB data, we use the three-year WMAP (WMAP3) temperature-temperature
and temperature-polarization power spectrum with the routine for computing the
likelihood supplied by the WMAP team \cite{wmap3:2006:1}. The LSS information
we use consists of the galaxy power spectrum from the Sloan Digital Sky Survey
(SDSS-gal; \cite{Tegmark:2003uf}); the luminous red galaxy power spectrum,
also from the SDSS (SDSS-lrg; \cite{Tegmark:2006az}) and the galaxy power spectrum
from the 2dF Galaxy Redshift Survey (2dFGRS-gal; \cite{Cole:2005sx}). To be
conservative, for all galaxy data we have used only the information at
$k\leq 0.1\hmpcinv$, corresponding to the linear regime. For SN Ia data, we
use two datasets: Riess sample of 182 SNe \cite{sn182} and ESSENCE sample of
192 SNe \cite{Miknaitis:2007jd,Davis:2007na}. We consider two dataset combinations:

\begin{description}
\item [(I) ] Riess-182+WMAP3+SDSS-gal+2dFGRS-gal;

\item[(II)] ESSENCE-192+WMAP3+SDSS-lrg+2dFGRS-gal.
\end{description}

In addition, for both datasets we also use the following
information: Hubble Key Project measurement
$H_{0}=72\pm8$~km~s$^{-1}$~Mpc$^{-1}(1\sigma)$ \cite{Hubble}; baryon
density information from the Big Bang Nucleosynthesis
$\Omega_{b}h^{2}=0.022\pm0.002$ ($1\sigma$) \cite{BBN}; and a
top-hat prior on the age of the universe, 10 Gyr $< t_0 <$ 20 Gyr.

For the MCMC calculation, we run eight independent chains each
originally with $\mathcal{O}(10^{5})$ elements, then thinned by a
factor of $4$. The average acceptance rate is about $50\%$. We
ensure the convergence of the chains by Gelman and Rubin criteria
\cite{R-1} and find $R-1\sim\mathcal{O}(10^{-2})$ which is more
conservative than the recommended value
$R-1\sim\mathcal{O}(10^{-1})$.

\begin{figure*}
\begin{center}
\hspace{-1.25cm}
\includegraphics[scale=0.75]{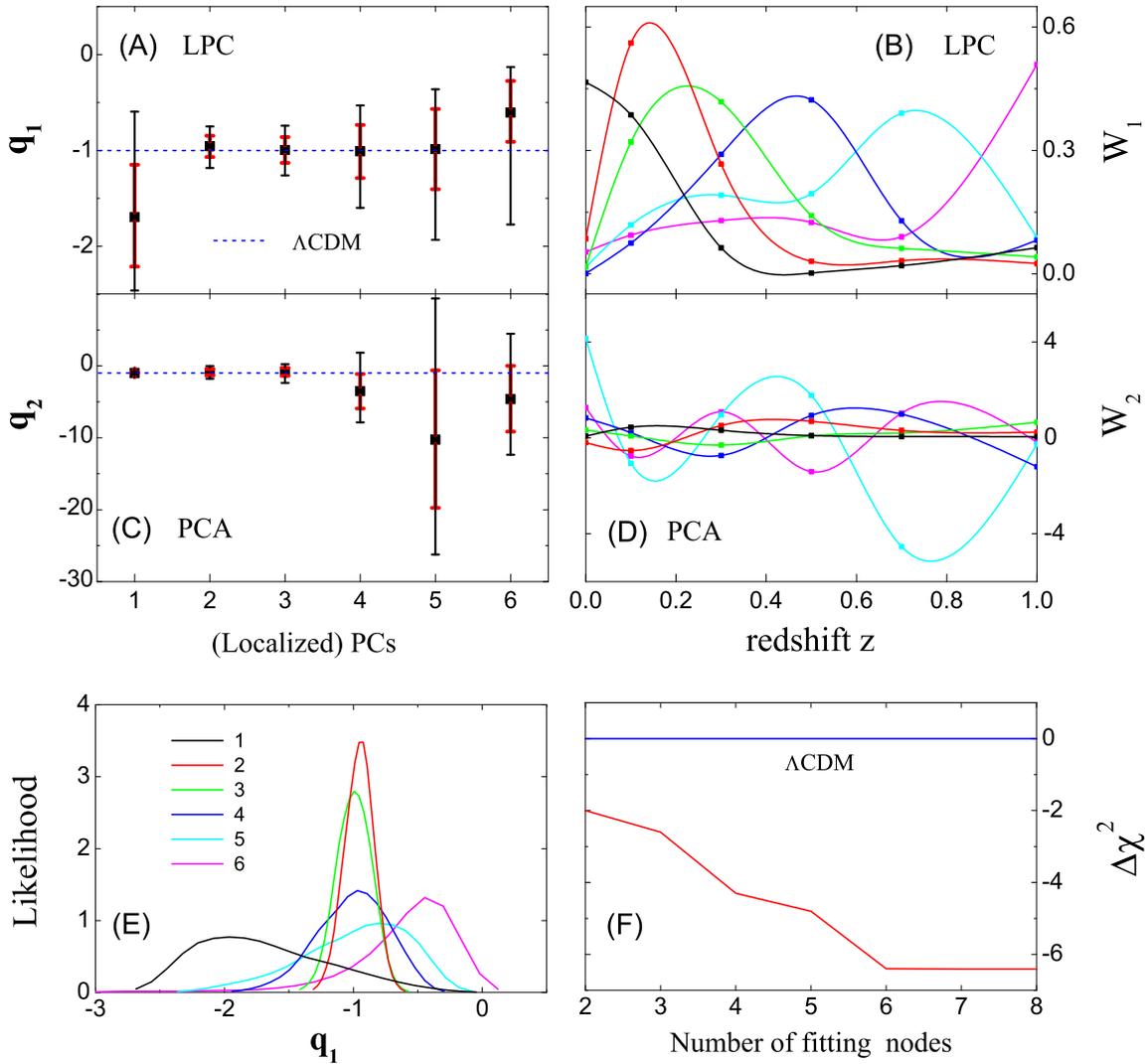}
\caption{Constraints on uncorrelated band powers of the equation of state
  $w(z)$, using the dataset II.  Panels (A) and (B) show the median
  values and the 68\% and 95\% constraints on the localized principal
  components $\tq_1$, and their weights in redshift $\tW_1$,
  respectively. Panels (C) and (D) show the same for the principal components
  $\tq_2$ and their weights $\tW_2$ (see the text for definitions).  Panel (E)
  shows the likelihood distributions for the parameters $\tq_1$, while panel
  (F) shows the reduction in $\chi^{2}$ relative to $\Lambda$CDM as the number
  of fitting nodes (for parameters $\tq_1$) is increased.}
\label{fig2}
\end{center}
\end{figure*}

\emph{Results}. We first study how the results change as we vary the
number of and location of fitting nodes describing $w(z)$.  We start
out by placing one fitting node each at $z=0$ and $z=\zmax=1$, then
increase $n$, adding fitting nodes so that they are uniformly spaced
in redshift; $z_i=(i-1)/(n-1), (n\geq2,~n\in\mathbb{N})$.  For each
$n$ and using the dataset (II), we calculate the improvement in
$\chi^{2}$ relative to the
  $\Lambda$CDM model, $\Delta\chi^{2}=\chi^{2}_{\rm min}-\chi^{2}_{\rm \Lambda
    CDM}$. As expected, we find that $n=2$ already has a better fit than the
$\Lambda$CDM model, and the fit improves as $n$ is increased.  When
$n$ is greater than $6$, however, $\Delta\chi^{2}$ flattens off (see
panel
  (F) of Fig.~\ref{fig2}). We therefore conclude that the data we use have
enough power to constrain $6$ equation of state parameters.  Therefore, at
least when $n\leq 8$ is considered, $n=6$ is the optimal choice since a higher
$n$ does not improve the fit\footnote{We are aware of that the fit must
  further improve for some higher $n$ because {\it any} dataset can be
  perfectly fit by some suitable, but possibly highly oscillatory, $w(z)$.}.

We also test the prescription of how to place the fitting nodes in redshift in
the lower panel of Fig.~\ref{fig1}. The black points with error bars are for
placement $z_i\in \{0,0.1,0.3,0.5,0.7,1.0\}$; the red points with red curve
are for placement $z_i\in \{0,0.2,0.4,0.6,0.8,1.0\}$. We see that the red
curve almost coincides with the black points, namely, the dependence of result
on placement of fitting nodes is weak. The above test and analysis has been
also done with dataset (I) and a similar result is found.  In what follows,
we adopt the former, slightly non-uniform placement in redshift.

Given the above two cosmological datasets, we explore the 12-D
parameter space from Eq.~(\ref{parameter}). After marginalizing over
other cosmological parameters, we obtain the constraint of DE
parameters $w_{i}$ as plotted in Fig.~\ref{fig1}. In the upper
panel, we find that almost all the mean values of points are
consistent with the prediction of $\Lambda$CDM model except for the
$1$st and the $4$th points, at redshifts $0$ and $0.5$. The equation
of state at the present epoch is slightly favored to be below $-1$,
while the equation of state at redshift $0.5$ is greater than $-1$
at 68$\%$ C.L., showing a small ``bump''. This bump most likely stems
from a feature in the Hubble diagram of the Riess-182 dataset (I).
To see this, we replace the Riess-182 and SDSS-gal data by
ESSENCE-192 and SDSS-lrg data respectively and get the result in
lower panel of Fig.~\ref{fig1}.  With this new dataset, the bump at
redshift $0.5$ disappears and $w(0.1\lesssim z\lesssim 0.9)$ is
consistent with $-1$ at $68\%$~C.L.. The equation of state at
redshifts $0$ and $1$ deviates from $-1$, but not significantly.

The equation of state parameters $w_i$ are correlated, however, somewhat
complicating their interpretation; for example, the highest correlation
coefficient is $-0.9$,  between $w_4$ and $w_5$.  It is therefore useful to
de-correlate the $w_i$, as done in Ref.~\cite{Huterer:2004ch}.

For each MCMC run, we compute the covariance matrix of the equation-of-state
parameters, $\textbf{C}=(w_i-\langle w_i\rangle)(w_j-\langle
w_j\rangle)^T\equiv \langle \tp \tp^T\rangle$, using \texttt{CosmoMC}
\cite{cosmomc}. We then diagonalize the Fisher matrix $\textbf{F}\equiv
\textbf{C}^{-1}$, so that $\textbf{F}=\textbf{O}^{T}\,\textbf{D}\,\textbf{O}$;
here $\textbf{O}$ is the resulting orthogonal matrix and $\textbf{D}$ is diagonal.

One can now rotate the parameters into a basis where the new parameters $\tq$
are uncorrelated; $\tq={\bf Wp}$. There are many ways to do so
\cite{Hamilton_Tegmark}. Here we make two alternative choices.

\begin{description}

\item[Choice 1:] $\tW_1=\tF^{1/2}$; then\\
$\tC_{\tq_1}=\langle \tq_1\, \tq_1^T\rangle = \tW_1\,\langle \tp
\tp^T\rangle\, \tW_1^T = \tI $.

\item[Choice 2:] $\tW_2=\tO$; then\\
$\tC_{\tq_2}=\langle \tq_2\, \tq_2^T\rangle = \tW_2\,\langle \tp
\tp^T\rangle\, \tW_2^T = \tD^{-1} $.

\end{description}

Here $\tW_1$ corresponds to the localized principal component
(heretofore LPC) decomposition of the parameters $\tp$ --- the
weights (rows of $\tW_1$) are almost positive definite and fairly
well localized in redshift \cite{Huterer:2004ch}. $\tW_1$ is usually
re-scaled so that its rows sum up to unity,
$\sum_{j=1}^{n}(\tW_1)_{ij}=1$, and we follow this practice. Note,
however, that the physical inferences do not depend on the
normalization of $\tW_1$ (or $\tW_2$) as both the parameters $\tq$
and their values for a particular theoretical model change
consistently.  For example, we have tried a different normalization
of $\tW_1$ and explicitly checked that the parameters $\tq$ and
their values corresponding to the $w(z)=-1$ model change so that the
hypothesis test of whether the measured parameters are consistent
with $w(z)=-1$ returns the same result.

We find that all LPC band powers are consistent with the cosmological constant
value at $95\%$ C.L. Two out of six points deviate from $\Lambda$CDM at
greater than $68\%$ C.L, which is not statistically unexpected. This
conclusion is easy to read off in the panel (A) of Fig.~\ref{fig2}
where the $\tq$'s are uncorrelated, and is independent of the choice of the
decorrelating weights. In panel (C), we replace $\tq_1$ with
$\tq_2$ (with $\tW_2$ normalized as $\tW_1$ above), and again find consistency
with the cosmological constant scenario.

\emph{Summary and Discussion.} Using the latest astronomical data of SNe Ia,
CMB and LSS, we used the MCMC machinery to reconstruct the expansion history
of the universe, and the evolution of dark energy, in a nearly
model-independent fashion.  We used the cubic spline interpolation between $n$
values of the equation state $w_i$, fitting the expansion history more
flexibly than essentially any description used on existing data so far.  We
found that $n=6$ values of the equation of state in the range $0\leq z\leq 1$
can be usefully constrained, and that increasing their number slightly does
not improve the fit.

Some comments can be made about our approach. We think the cubic spline leads
a largely unbiased reconstruction of $w(z)$. An exception to this would be a
highly oscillatory fiducial $w(z)$, which however cannot be robustly
constrained with other approaches (and current data) either.  Moreover, our
results are not in conflict with Ref.~\cite{howmany} who claimed that only 2-3
equation-of-state parameters can be measured even from future surveys to
better than about 10\%, since we do not measure the six $w_i$ to nearly such a
good accuracy; see Fig.~\ref{fig1}. Finally, one could certainly apply the
same approach to the reconstruction of the dark energy density $\rho_{\rm
  DE}(z)$ as in \cite{Wang_Mukherjee}, although we would argue that probing
dynamics of dark energy requires specifically $w(z)$.

Our results do not show any significant evidence for the evolution
of the equation of state with time, and are fully consistent with
the cosmological constant scenario in the interval $0\leq z\leq 1$
at the 95\% C.L.  We would particularly like to be able to test the
dynamics of dark energy; for example, the ``freezing'' and
``thawing'' scalar field models \cite{Caldwell_Linder} which have
different physical behavior and a different sign of $dw/dz$; however
the accuracy we have right now is not sufficient to distinguish
between these models \cite{Huterer_Peiris}.  Our method is
straightforward to implement, and will produce sharp tests of the
dynamics of dark energy once we have data from the next-generation
surveys, such as Dark Energy Survey \cite{des}, Joint Dark Energy
Mission \cite{jdem}, Large Synoptic Sky Telescope \cite{lsst}, and
Pan-STARRs \cite{pan}.

\vspace{-0.3cm}
\begin{acknowledgments}
\vspace{-0.35cm}

We acknowledge the use of the Legacy Archive for Microwave
Background Data Analysis (LAMBDA) provided by the NASA Office of
Space Science. All of our numerical analysis are performed on the
Shanghai Supercomputer Center (SSC). We thank Levon Pogosian, Andrei
Frolov, Jun-Qing Xia and Alireza Hojjati for helpful discussions. DH
thanks the Institute for High Energy Physics and the National
Astronomical Observatories in Beijing for hospitality. This work is
supported in part by National Natural Science Foundation of China.
G.Z. is supported by National Science and Engineering Research
Council of Canada (NSERC).
\end{acknowledgments}

\end{document}